\documentclass[jcp,aps,twocolumn,a4paper]{revtex4}

\usepackage{color}
\usepackage[sort&compress]{natbib}
\usepackage{amssymb}
\usepackage[T1]{fontenc}
\usepackage[ngerman,english]{babel}
\usepackage{amsfonts}
\usepackage{amsmath}
\usepackage{array}
\usepackage{epsf}
\usepackage{epsfig}
\usepackage{float}
\usepackage[version=3]{mhchem}
\usepackage{textcomp}
\usepackage{graphics}
\usepackage{graphicx}
\usepackage{epstopdf}
\usepackage{placeins}
\usepackage{footnote}
\usepackage[para]{threeparttablex}
\makesavenoteenv{tabular}
\makesavenoteenv{table}
\newcolumntype{x}[1]{>{\centering\arraybackslash\hspace{0pt}}p{#1}}

\begin{document}
\newcommand{\Fig}[1]{Fig.~\ref{#1}}
\newcommand{\eq}[1]{Eq.~(\ref{#1})}
\newcommand{\tc}[1]{\textcolor{red}{#1}}

\author{Olga~S.~Bokareva}
\email{olga.bokareva@uni-rostock.de}
\author{Oliver~K\"{u}hn}

\affiliation{
Institut f\"ur Physik, Universit\"at Rostock, Universit\"atsplatz 3, D-18055 Rostock, Germany\\
}

\date{\today}

\title{Quantum chemical study of the electronic properties of an Iridium-based photosensitizer bound to medium-sized silver clusters}

\begin{abstract}
The equilibrium structures and electronic excitation spectra of the Ir(III) photosensitizer \ce{[Ir(ppy)2(bpy)]+}
bound to medium-sized silver clusters \ce{Ag_{$n$}} ($n=$19, 20)
are investigated using time-dependent density functional theory.
The long-range corrected LC-BLYP approach is used with a system-specific 
range separation parameter. 
The weak physisorption of the hybrid complexes yield only small changes in the 
broadened absorption spectra of the hybrid system as compared with its constituents. However, the density of states as well as the fine structure of the spectra is strongly modified upon complexation. It is shown that the standard range separation parameter (0.47 bohr$^{-1}$) cannot predict these properties correctly and the optimized value of 0.16 bohr$^{-1}$ should be used instead.
\end{abstract}

\maketitle
%
\section{Introduction}
%

The combination of metal nanoparticles or nanowires
with various organic adsorbates such as dyes, peptides, and J-aggregates is an active area of research with applications in bio-sensoring, catalysis, and medicine~\cite{Yu-acie-2007, Tanaka-acie-2011, Sapsford-an-2011,Halivni-an-2012, Tel-aemb-2012}. 
Numerous theoretical and 
experimental studies of  nanoparticle-organic hybrid
systems have been reported (for reviews, see, e.g., Refs.~\cite{Dulkeith-prl-2002, Anger-prl-2006, 
Pustovit-prb-2011, Zweigle-jpcc-2011}).
These hybrid systems exhibits new composition dependent properties, which differ from those of the separate constituents
This includes the 
enhancement or quenching of fluorescence, absorption,
and Raman scattering due to surface and plasmon resonances, the broadening of the absorption range to yield an antenna effect, and modified ``redox'' properties; see, e.g., 
Refs.~\cite{Wenseleers-jpcb-2002, Lee-nl-2005, Saini-jppac-2007, Zou-np-2012, Rai-ao-2012}.

\begin{figure}[t]
\includegraphics[width=0.3\textwidth]{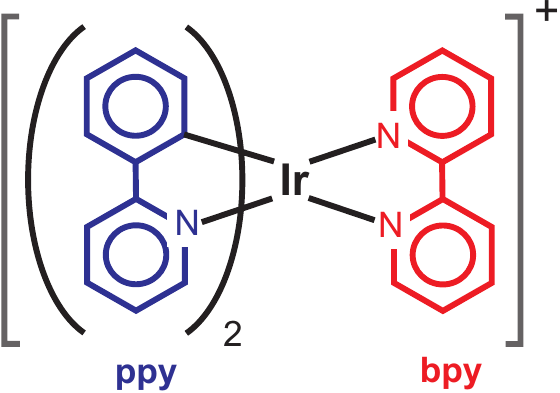}
\caption{\label{IrPSscheme}
Chemical formula of the photosensitizer \ce{[Ir(ppy)2(bpy)]+} (IrPS).}
\end{figure}

In the present communication we discuss results on the electronic properties of a model system comprised of the \ce{[Ir(ppy)2(bpy)]+} photosensitizer (IrPS)  shown in \Fig{IrPSscheme}, bound to 
medium-sized silver clusters \ce{Ag_{$n$}} ($n$=19, 20). 
Our choice of the system is motivated by the use of heteroleptic
Ir(III) complexes as photosensitizers in a  photocatalytic system for water splitting
introduced by Beller et al.\  \cite{Gaertner-acie-2009}.
The original homogeneous system includes triethylamine as a sacrificial reductant  
 and a series of iron carbonyls as water reduction
catalysts. The absorption spectrum of the IrPS overlaps with the sun's spectrum only in its long wave length tail between 300 and 450~nm~\cite{Bokarev-jcp-2012}.
Therefore, extending the range of absorption further  into the visible by coupling the IrPS to silver nanoparticles might provide a means to enhance the overall efficiency of this photocatalytic system.

Because of size of the systems under study,
density functional theory (DFT)
\cite{Koch2001}
and its extension into the  time-dependent (TD) domain in the linear response limit
\cite{Casida1995}
is the natural choice  for studying  ground and excited state properties.
However, the correct description of long-range
charge-transfer (CT) properties, which is of mandatory for 
hybrid systems, can not be achieved applying
conventional DFT functionals \cite{Dreuw-jcp-2003, Dreuw-cr-2005, Peach-jcp-2008, Richard-jctc-2011,bokarev15_xxx}.
By introducing exact Hartree-Fock exchange  in long-ranges corrected schemes
as LC-BLYP \cite {Iikura-jcp-2005, Tawada-jcp-2004, Chiba-jcp-2006} or CAM-B3LYP \cite{Yanai-cpl-2004},
the correct asymptotic behavior and more  balanced description for CT states is obtained.
In our previous study on bare IrPS \cite{Bokarev-jcp-2012},  the LC-BLYP approach was shown 
to give excitation energies in good agreement with experimental data and results of
CASSCF/CASPT2 calculations.

The binding of IrPS to small model silver clusters  (one to six silver atoms) has been studied using the LC-BLYP approach in Ref.~\cite{Bokareva-pccp-2012}.
Already for this size of the hybrid system pronounced changes in the electronic absorption spectrum had been observed. Moreover, the electronic properties were found to be rather sensitive to the number of silver atoms, Binding energies and localization of HOMO and LUMO orbitals are found to oscillate with the number of silver atoms in an even/odd like fashion. Whether such a  behavior propagates to larger clusters and how it influences the photophysical properties is the topic of the present study.

The paper is organized as follows. In Sec.~\ref{sec:Comp} we give the computational details putting emphasis on the tuning of the 
range-separation functional in LC-BLYP. The results are presented in Sec.~\ref{sec:Res}, including electronic ground state structures and electronic absorption spectra. A summary is provided in Sec. \ref{sec:Conc}.
%
\section{Computational details}
\label{sec:Comp}
%
In the long-range separation approach LC-BLYP a parameter $\omega$ is introduced,
which defines the  separation of the Coulomb operator into long-range and short-range parts, with the long-range part then being described by the exact exchange integral.
In a number of publications, it was demonstrated
that an appropriate tuning of $\omega$ leads to a significant improvement of fundamental  and optical gaps, CT and Rydberg excitation energies as well as ionization potentials (IPs)  \cite{Sears-jcp-2011, Korzdorfer-jcp-2011, Karolewski-jcp-2013, Refaely-prb-2011}. A systematic procedure for the determination of an optimal $\omega$ has been suggested in Refs.~\cite{Livshits-pccp-2007, Stein-jacs-2009, Stein-jcp-2009}. It is based on finding that $\omega$, which minimizes the following functional 
\begin{eqnarray} \label{Jsum} 
J(\omega ) &=& J_{N}(\omega )+ J_{N+1}(\omega) \nonumber\\
&=& \left|\varepsilon^{\omega }_{\rm HOMO}(N)+\text{IP}^{\omega}_{}(N)\right| \nonumber\\
&&+ \left|\varepsilon^{\omega }_{\rm HOMO}(N+1)+\text{IP}^{\omega}_{}(N+1)\right| \, .
\end{eqnarray}
In other words, the optimized $\omega$ will provide a compromise for fulfilling Koopmans' theorem simultaneously for systems with $N$ (cation complex) and  $N+1$ (neutral complex)  electrons. In \eq{Jsum},
$\varepsilon^{\omega }_{\rm HOMO}$ and $\text{IP}^{\omega}_{}$ is the HOMO energy
and the IP, respectively.

In Ref.~\cite{Bokareva-omega-2014} we have studied the present systems and obtained  optimized range-separation parameters for the bare IrPS, silver clusters \ce{Ag_{$n$}}, and hybrid systems \ce{IrPS-Ag_{$n$}} ($n$ = 2, 10, 20). Here, we give a more detailed account on the case of \ce{IrPS-Ag_{$10$}}. In short, 
the IPs and HOMO energies of neutral species, anions, and cations have been obtained from single-point
energy calculations at the optimized  geometry of the electronic ground state. 
Similar to the small systems \cite{Bokareva-pccp-2012},
the geometry optimization was carried out without
symmetry constraints. Initial geometries of IrPS 
for further optimization were taken from Ref.~ \cite{Bokarev-jcp-2012} 
(assuming $\text{C}_{2}^{}$ point symmetry \cite{King-jacs-1987}). The initial structures of \ce{Ag10} were taken from Ref.~\cite{Yang-jcp-2006}, those of \ce{Ag19} and \ce{Ag20} from Ref.~\cite{Baishya-prb-2008}, except tetrahedral  \ce{Ag20} which had been studied before in  Refs.~\cite{Aikens-jpcc-2008,  Zhao-nl-2006, Lang-chc-2011}.
Binding energies have been obtained including the 
counterpoise method to correct for the basis set
superposition error (BSSE) \cite{Boys-mp-1970, Simon-jcp-1996}. 

Vertical excitation spectra have been calculated
by means of the TDDFT approach. The number of included transitions  has been 450 and 650 for XXa and IXXo, respectively.
A Lorentzian broadening 
(0.4 eV) has been added to the  stick-spectra. Spin-forbidden transitions were not included in
the present TDDFT investigation although
they could have notable intensity due to the high spin-orbit coupling constant of Ir. 	
All calculations were performed with the Gaussian09 program package  \cite{g09}  using the LANL2DZ effective core potential  basis set for Ir and Ag
and the 6-31G(d) basis for all other atoms.

\begin{figure}[t]
\includegraphics[width=0.5\textwidth]{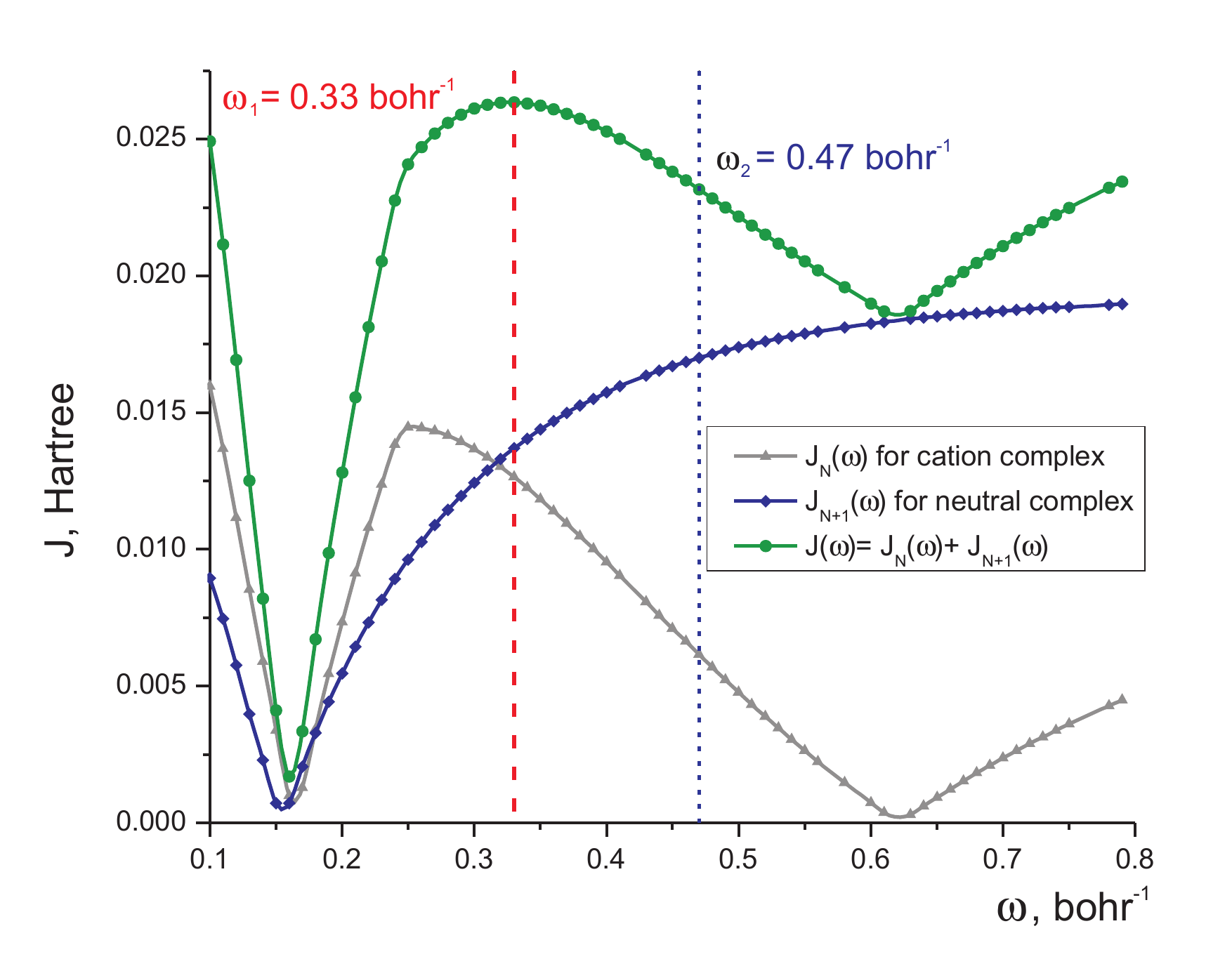}	
	\caption{\label{omegaX}
		Optimization of range-separation parameter for the case of \ce{IrPS-Ag_{10}^{}}. The functions $J^{}_{N}(\omega)$ and 
			$J^{}_{N+1}(\omega)$ are defined in \eq{Jsum} and represent the conditions for fulfilling Koopmans' theorem for the cationic and neutral  species, respectively.}
\end{figure}

%
\section{Results and discussion}
\label{sec:Res}
\subsection{Range-separation Parameter Optimization}
In \Fig{omegaX}, the procedure for tuning $\omega$
is demonstrated for the case of \ce{IrPS-Ag_{$10$}}.
A minimum of $J(\omega)$ at 0.16 $\text{bohr}_{}^{-1}$ was found.
Around 0.63 $\text{bohr}_{}^{-1}$
an  additional higher energetic minimum occurs. This can be traced to the change of the character of the lowest adiabatic state of the cationic species, where the order of  HOMO and HOMO-1 orbitals are interchanged. Here, the 
unpaired electron initially localized on the $\pi^{*}_{}(\text{bpy})$ orbital of the reduced doublet IrPS
 is transferred  to the $\sigma(\text{Ag}_{10}^{})$ orbital~\cite{Bokarev-pccp-2014}.

The obtained optimal $\omega$ value is substantially
smaller than values of $0.33 \, \text{bohr}_{}^{-1}$
and $0.47 \, \text{bohr}_{}^{-1}$ 
(shown as vertical lines in \Fig{omegaX})  implemented in common quantum chemical packages. This is not surprising since these  (standard) values were determined for test sets  of diatomics and small molecules \cite{Tawada-jcp-2004, Song-jcp-2007}. 

Since the inverse of $\omega$ reflects a characteristic 
distance for switching between short- and 
long-range parts of the exchange contribution, optimal
$\omega$ values were shown to decrease with increasing system size and 
conjugation length \cite{Stein-jcp-2009, 
Korzdorfer-jcp-2011, Stein-prl-2010, 
Karolewski-jcp-2011, Refaely-prb-2011, 
Sears-jcp-2011, Salzner-jctc-2011}. 
This size-dependence, however, is not monotonous 
and there is a strong dependence on the electronic 
structure \cite{Refaely-prb-2011}. Optimal values
of $\omega$ near $0.2 \: \text{bohr}^{-1}_{}$
have been reported for various systems 
with  characteristic electronic radii
comparable to the present ones, e.g., 0.214 $\text{bohr}^{-1}_{}$ was found 
for pentacene-\ce{C60} \cite{Minami-ijqc-2013}; for other
examples see Refs.~\cite{Stein-jcp-2009, Stein-prl-2010, Refaely-prb-2011}. 

As shown in Ref.~\cite{Bokareva-omega-2014} there is only a slight variation of $\omega$ when going from IrPS, via Ag$_n$ ($n=10,20$) to \ce{IrPS-Ag_{$10,20$}}.  
Thus, in the following we provide results on excited states obtained
using  $\omega=0.16 \, \text{bohr}_{}^{-1}$ for all compounds. 
It turned out the ground state geometries of the present systems are little affect by the choice of $\omega$.  The corresponding tests have been performed for IrPS, \ce{Ag2}, \ce{IrPS-Ag2}, and \ce{IrPS-Ag19}.
Therefore, we have used the 
the standard $\omega^{}_{2}=0.47 \, \text{bohr}_{}^{-1}$ for the ground state optimization. Selected TDDFT calculations are also performed with the latter value in order to study the  influence of $\omega$ on the absorption spectrum.

\subsection{Ground electronic state}
%

\begin{figure}[t]
\includegraphics[width=0.45\textwidth]{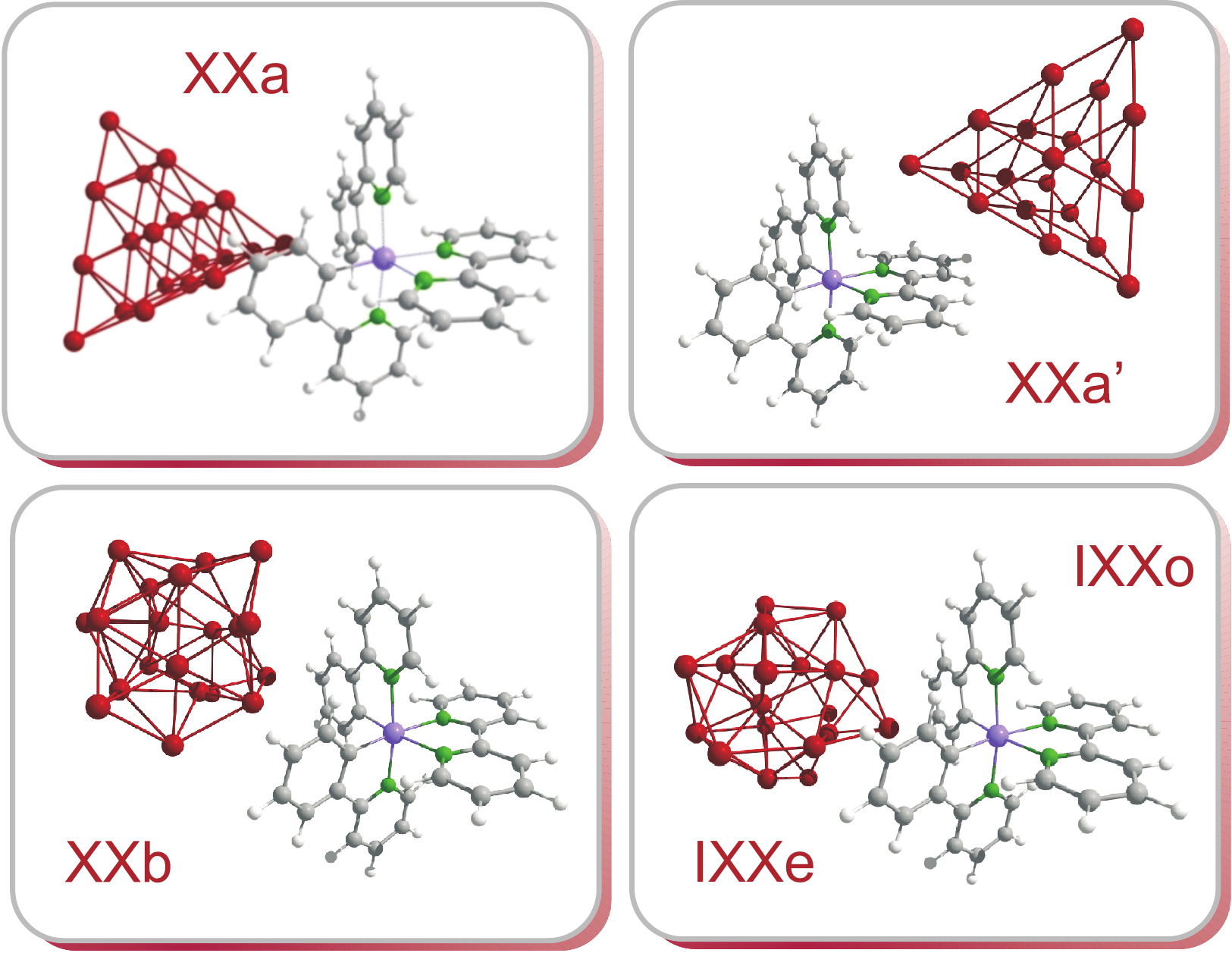}
\caption{\label{Structures}
Optimized structures of \ce{IrPS-Ag19} and
\ce{IrPS-Ag20} studied in this paper. The conformers labeled IXXe and IXXo are not distinguishable at the scale of the image.}
\end{figure}

In our previous study of IrPS bound to small ($n\le 6$)
silver clusters \cite{Bokareva-pccp-2012}, we found
that configurations in which \ce{Ag_{$n$}} is 
situated in the cavities
between ligands are the lowest in energy. Note
that in all cases the interactions are ``weak'' and
no chemical bonds are formed. Here, we extended 
our investigation to \ce{IrPS-Ag_{$n$}} geometries with 19 and 
20 silver atoms, however, focussing mostly on 
those structures where the cluster is located  in the ``ppy-ppy'' cavity
as shown in \Fig{Structures}.  

The differences in geometry of \ce{IrPS} upon complexation with a silver cluster are only minor (maximal changes are 0.09 \AA{} and $4.6^{\circ}_{}$); for an overview of geometries see Table ~\ref{Geom}.
The binding energies of the  studied systems
are in the range $1.5-2.5\: \text{kJ/mol}$, which is about 10 times smaller than that of silver atoms in large nanoparticles \cite{Fernandez-prb-2004} and smaller than in analogous systems with $n=1-6$. 
This confirms the trend of a decrease of binding energies with the increase of cluster size found previously \cite{Bokareva-pccp-2012}.
Such a behavior can be explained by the fact
that no stable complexes are formed, i.e. the interaction
belongs to physisorption case and might
be considered as a weak attraction between 
the ligands of IrPS and nearest atoms of silver cluster.
The total  binding energies 
(not divided by number of silver atoms)
are 30-100~kJ/mol for all systems disregarding the cluster size.
Accounting for the BSSE decreases the binding energy by factor of  about two for all cases. The change of $\omega$
from the standard to the optimized value has no uniform impact on
binding energies. Changes of a  factor of about two in both directions were observed.
Note that the order of stability is sensitive to the value of range-separation parameter used. For clarity, we provide here only results of calculations with the optimal $\omega$ value. For the series of XX stuctures, the XXa conformer is the lowest in energy, while XXa' and XXb are by 30.0 and 4.1 kJ/mol higher, respectively. For the IXX species, IXXe is more stable ($\Delta \text{E} = 38.2 \text{kJ/mol}$). Nevertheless, in the following we also discuss the IXXo structure because of its unique electronic structure in the ground electronic state.

\begin{table*}
\begin{threeparttable}
\caption{\label{Geom}
	Selected equilibrium geometric parameters of all 
	\ce{IrPS}$\cdots$\ce{Ag}$_{n}^{}$ compared to IrPS \cite{Bokarev-jcp-2012} and \ce{IrPS}$^{\text{0}}_{}$ \cite{Bokarev-pccp-2014} in the ground electronic states.
	For notation of structures see \Fig{Structures}.
	Bond lengths in \AA, angles in degrees.}
\begin{tabular*}{0.9\textwidth}{@{\extracolsep{\fill}} l c c c c c c c}
\hline
Parameter \tnote{a} & IrPS & IrPS$^{\text{0}}$ &
 IXXo & IXXe & XXa & XXa$^{\prime}_{}$ & XXb \\
\hline
 r(Ir--C$_{\text{ppy}}^{})$ & 2.007 & 2.010 &  2.019 & 2.005 & 2.007 & 2.001 & 2.002 \\
 r(Ir--N$_{\text{ppy}}^{})$ & 2.055 & 2.051 & 2.051 & 2.059 & 2.058 & 2.056 & 2.057 \\   
 r(Ir--N$_{\text{bpy}}^{})$ & 2.163 & 2.134 & 2.116 & 2.158 & 2.158 & 2.160 & 2.162 \\  
 r(C--C)$_{\text{ppy}}^{}$ & 1.463 & 2.464 & 1.470 & 1.463 & 1.466 & 1.457 & 1.461 \\ 
 r(C--C)$_{\text{bpy}}^{}$ & 1.485 & 2.409 & 1.408 & 1.482 & 1.485 & 1.481 & 1.482 \\ 
 r(Ir--Ag) \tnote{b} & -- & -- & 3.563 & 3.919 & 4.042 & 5.101 & 5.108 \\   
 $\angle$CIrC & 89.3 & 89.1 & 93.9 & 92.9 & 90.7 & 89.8 & 90.4 \\   
 $\angle$N$_{\text{ppy}}^{}$IrN$_{\text{ppy}}^{}$ & 172.1 & 174.5 & 174.8 & 171.8 & 171.8 & 172.8 & 172.2 \\
  $d$CIrCN & 92.4 & 96.1 & 95.8 & 93.2 & 93.3 & 95.0 & 94.6 \\                                                                                      
 \hline
\end{tabular*}
\begin{tablenotes}
	\item[a] For parameters including ppy, only those ppy fragment that is closer to the silver cluster is regarded. The parameters of second ppy fragment differ by no more than 0.04~$\AA$;
	\item[b] Distance between Ir and closest silver atom.
\end{tablenotes}
\end{threeparttable}
\end{table*}

\begin{figure}[t]
\includegraphics[width=0.49\textwidth]
	{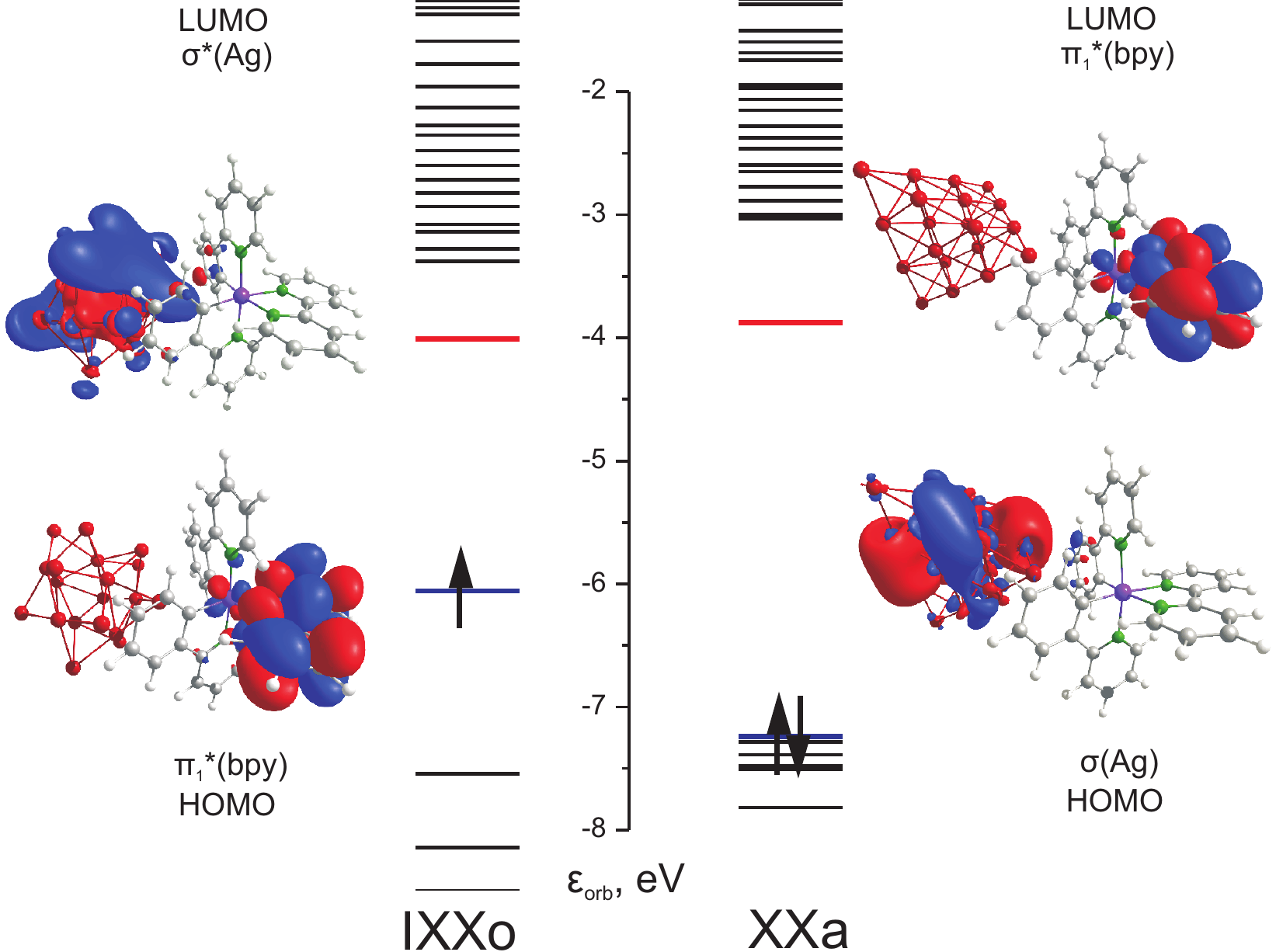}
\caption{\label{Orbitals}
	Molecular orbital energies and frontier orbitals for IXXo and XXa systems.}
\end{figure}

The representative frontier orbitals (LC-BLYP, $\omega=0.16\: \text{bohr}^{-1}_{}$) are given in \Fig{Orbitals} for the cases of IXXo and XXa.
If the standard value of $\omega$ is applied then the HOMO-LUMO gap increases by about 1 eV for all cases. The shift is mainly due to the LUMO orbitals while the nature and order of orbitals remains the same. 

The orbital structure differs for systems containing odd and even number of silver atoms. 
For an even number, the HOMO orbital is localized on the silver fragment ($\sigma ^{}_{\text{Ag}}$) and the 
LUMO is of $\pi^{*}_{1}\text{(bpy)}$ type.
For an odd number of silver atoms this situation is reversed:
 the unpaired electron is situated on the  $\pi^{*}_{1}\text{(bpy)}$ orbital as in the reduced IrPS, while the LUMO is localized on the  silver fragment. 
A similar situation with one exception for \ce{IrPS-Ag5} was found for smaller clusters \cite{Bokareva-pccp-2012} as well as for the much larger cluster \ce{IrPS-Ag92} \cite{Bokareva-cp-2014}.

For 19 atoms  two cases were found which slightly differ  in geometry: IXXo (``o'' means ``odd'') follows the orbital rules just given and IXXe (``e'' stands for ``even'') is an exception where the unpaired electron is localized on the silver fragment. 
The  ``odd'' and ``even'' groups can be also seen from the comparison
of geometries, see Table ~\ref{Geom}; the distances Ir-N$_{\rm bpy}^{}$ and $r$(C-C)$_{\rm bpy}^{}$ are shorter for the ``odd'' group and the IrPS in its reduced form (IrPS$^0$).


\subsection{Absorption spectra}
\begin{figure*}[t]
\begin{widetext}
\includegraphics[width=0.8\textwidth]{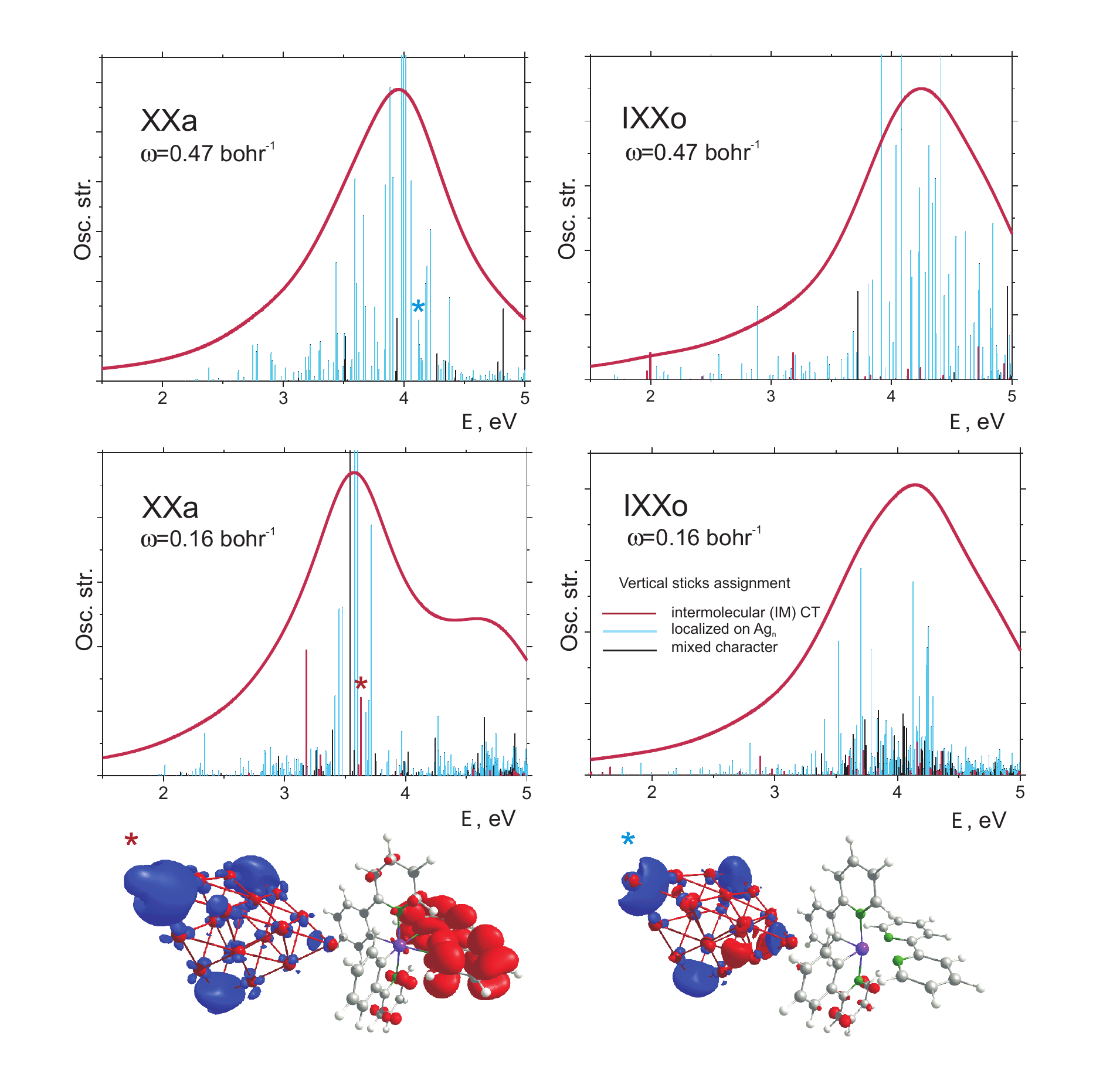}
\caption{\label{Spectra_all}
Upper panel: Selected vertical electronic singlet-singlet (for XXa) and 
doublet-doublet (for IXXo) TDDFT spectra 
calculated with LC-BLYP with standard and optimized $\omega$ parameter.
Associated spectra are broadened with Lorentzians of width 0.4 eV. 
Lower panel: Electron density differences (contour value 0.0008) for selected transitions, 
which are marked with asterisks in upper panel.  Red and blue colours correspond to particle and hole densities, respectively.}
\end{widetext}
\end{figure*}

The absorption spectra of IXXo and XXa are presented 
in \Fig{Spectra_all}.  The standard (0.47 bohr$^{-1}_{}$) and optimized
$\omega$ values were applied for comparison. 
Because of the number of states considered and of
the complex character of most states, we distinguish  only between
two types of orbitals: those localized on silver or on dye fragments, but 
do not consider the particular character of orbitals, e.g., $\pi\rm{(ppy)}$, 
$\pi\rm{(bpy)}$, or $d\rm{(Ir)}$ for IrPS. For each state different contributions have
been summed up and the corresponding dominating character has
been assigned. Most of transitions have very complex assignment and
are marked as ``mixed''.
In \Fig{Spectra_all} the spectra are shown as sticks and with a Lorentzian broadening.

All spectra have a clear  maximum composed mainly by
silver-localized transitions, which, for pure cluster spectra, would  be called ``plasmonic'' excitations (collective excitation of electrons).
Similar to the cases of small silver clusters \cite{Bokareva-pccp-2012}, the hybrid systems support a  new type of long-range intermolecular CT, corresponding to
$\rm{IrPS}\rightarrow \rm{Ag_{n}^{}}$ or $\rm Ag_{n}^{} \rightarrow \rm IrPS$. These
intermolecular CT transitions (denoted as IM to
emphasize the their  long-range nature) are marked with red bars in \Fig{Spectra_all}.
The IM transitions have very different intensities, from almost zero to that comparable with silver plasmon-like transitions.
Many transitions can not be attributed to pure IM type. but have a notable contribution of  intramolecular CT character. These mixed transitions are marked with black bars in \Fig{Spectra_all}.
In the lower panel of \Fig{Spectra_all}, electron density difference plots for transitions of IM and cluster localized types are given. 
 
Comparing the spectra of XXa and IXXo with different $\omega$ values, three important points should be highlighted.
First, their maxima  are shifted to the red by 0.4 eV  for XXa and by 0.1 eV for IXXo.
The positions of particular bands can hardly be compared
as the assignment is very complex.
Second, the density of states for the optimized $\omega$
is almost twice higher than for the standard $\omega$. Finally, the distribution of oscillator strength depends on $\omega$ as well. This is particularly visible for the XXa system, where the optimized $\omega$ yields a double peak spectrum, which is in contrast to IXXo. For the standard $\omega$ XXa and IXXo have essentially the same line shape.
 
 
\begin{figure}[t]
\includegraphics[width=0.45\textwidth]{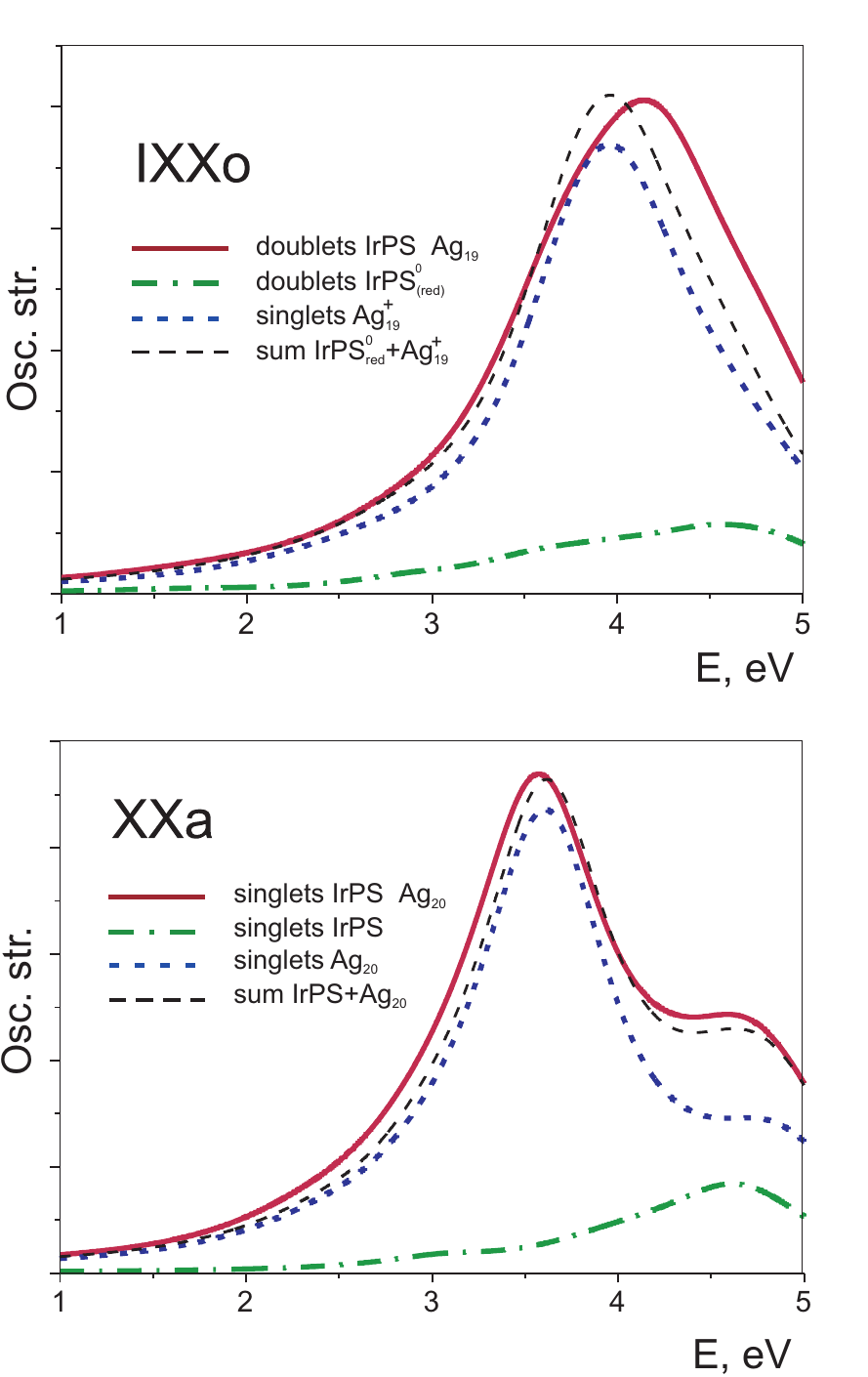}
\caption{\label{Comparison}
Calculated (LC-BLYP, $\omega = 0.16 \, \text{bohr}^{-1}_{}$) absorption spectra of pure IrPS, \ce{Ag_{n}}, their sum,  and combined systems \ce{IrPS-Ag_{$n$}} ($n$ = 19, 20). The width of the Lorentzian broadening has been 0.4 eV. Note that for IXXo the reduced IrPS and the oxidised Ag$_{19}$ have been used as reference, in accord with the ground state orbital structure of this system.
}
\end{figure}

The changes of absorption spectra upon complexation are of general importance for possible photochemical and photophysical applications. 
Because of complicated assignment of single bands, the direct comparison of particular transitions is hindered and only the overall shapes can be compared. In \Fig{Comparison} the broadened spectra for IXXo and XXa are compared to those of the pure constituents.
For IXXo, we provide the spectra of reduced \ce{IrPS^{0}} and Ag$_{19}^+$ according to the ground electronic state structure, whereas for XXa  the results for IrPS and \ce{Ag20} are given.

The broadening parameter was chosen in such a way as to reproduce
the approximate width of experimental plasmonic band for silver clusters, which is about 50 nm \cite{Thomas-n-2008, Ma-apa-2011}. However, 
different types of transitions (local and CT), even in
case of pure IrPS, could have a different broadening, which is not considered due to the lack of detailed information. 
For the case of smeared fine structures of spectra, we expect
our results to be only slightly dependent on the variation
of line shape and  small changes in line widths of transitions of particular types.
When discussing the changes in the broadened spectra in the following, one should keep in mind that the individual  transitions and thus the  unresolved fine structure of the spectrum is strongly modified upon complexation.

If we compare the absorption spectra of the hybrid systems with those of their constituents the following observations can be made: The maximum of the plasmonic band in the spectrum of IXXo is shifted by 0.2 eV to the blue and its width increases slightly.
For XXa the maximum remains almost at the same position upon complexation with IrPS. The position and shape of this broad band is very similar for unbound and bound clusters.
Some additional features in the spectra of 
XXa in the region from 4.5 to 5 eV could be attributed to the influence of IrPS through the manifold of new states with non-negligible IM CT character.

Figure \ref{Comparison} also contains the  sum of spectra of the isolated dye and the cluster. It  almost does not differ from the spectrum of the interacting hybrid system for  XXa, and it is shifted only by about 0.2 eV to the blue for IXXo.
But the density of states increases and most transitions have at least an admixture of IM character. Therefore, the spectra of hybrid systems cannot be considered as perturbed spectra of metal clusters.
Consequently, the photodynamics of the hybrid systems might strongly differ from those of pure IrPS.

%
\section{Conclusions}
\label{sec:Conc}
The interaction of an Ir(III)-based photosensitizer with silver clusters containing 19 or 20 atoms has been studied by means of TDDFT with a properly chosen range-separation parameter in the LC-BLYP functional.  Although the interaction between dye and metal particle belongs to the case of weak physisorption, the properties of the hybrid systems differ notably from those of its constituents. This concerns 
the appearance of long-range intermolecular CT states, which might lead to a completely different photophysical and photochemical behaviour, including long-lived charge separation favourable for further reactions. The low resolution electronic excitation spectra are close to the sum of the spectra of the non-interacting constituents.  However, the underlying density of electronic states and thus the fine structure of the spectra is strongly modified in the hybrid system, with the details depending on the number of silver atoms.
The changes in the  adsorption spectrum  upon forming the hybrid system are mostly observed the  region from 3 to 4.5 eV, but they are masked by the  broad plasmonic type absorption band of the cluster. 

If we extrapolate the obtained theoretical results to experimental observations the following considerations should be taken into account. 
First, metal clusters used for spectroscopic and catalytic studies are produced either chemically by reduction in solutions~\cite{Badr-ass-2006, Peng-apa-2012} or with laser techniques, e.g., cluster beam generation utilizing arc  discharge, magnetron sputtering, or laser vaporization 
(for review see, e.g., Ref.~\cite{Popok-ssr-2011}). 
The resulting clusters are substantially larger compared to those
used in our theoretical model and have a certain size distributions.
Second, the clusters are produced either in solution or deposited on a plate. 
The deposited nanoclusters have the advantage of being protected from fast aggregation and thus are often applied in spectroscopic studies and also in catalysis providing more stable and long-living catalysts. 
The experimental maximum of plasmonic band for silver particles is in the range 375-430 nm for an average diameter  from 2 to 50 nm \cite{Peng-apa-2012, Badr-ass-2006,
Hilger-apb-2001, Ma-apa-2011,Thomas-n-2008} and produced in aqueous solution or embedded in solid substrate (\ce{SiO2}, \ce{Cr2O3}, and \ce{MgF2}).
The maximum of this peak shifts to the red with increasing the average cluster  size and with increasing the dielectric constant of the surrounding medium. 
Consequently, the spectra of IrPS bound to silver clusters relevant in those experiments should be averaged over different cluster sizes and red-shifted as  compared to those shown in \Fig{Comparison}. It can be expected that the differences between odd and even number of atoms in cluster will not be seen in this case. Further, the interaction between the silver clusters and the IrPS will be weakened due to the increase of the spectral separation. The consequences for the properties of the hybrid system still need to be explored.

\acknowledgments
This work has been supported by the European Union (European Social Funds, ESF) within the project ''PS4H'' and by the Ministry for education, science and culture of Mecklenburg-Vorpommern.


\end{document}